\documentclass[journal=jctcce,manuscript=article,layout=twocolumn]{achemso}
\setkeys{acs}{articletitle = true}

\usepackage[T1]{fontenc}
\usepackage[utf8]{inputenc}
\usepackage{graphicx}
\usepackage{amsmath}
\usepackage{color}
\usepackage{comment}
\usepackage{hyperref}
\usepackage{multirow}
\usepackage{subfig}
\usepackage{braket}
\usepackage{bm}
\usepackage{widetext}

\title{Massively parallel quantum chemical density matrix renormalization group method}

\author{Ji\v{r}\'{i} Brabec}
\affiliation{J. Heyrovsk\'{y} Institute of Physical Chemistry, Academy of Sciences of the Czech \mbox{Republic, v.v.i.}, Dolej\v{s}kova 3, 18223 Prague 8, Czech Republic}

\author{Jan Brandejs}
\affiliation{J. Heyrovsk\'{y} Institute of Physical Chemistry, Academy of Sciences of the Czech \mbox{Republic, v.v.i.}, Dolej\v{s}kova 3, 18223 Prague 8, Czech Republic}
\alsoaffiliation{Faculty of Mathematics and Physics, Charles University, Prague, Czech Republic}

\author{Karol Kowalski}
\affiliation{Pacific Northwest National Laboratory, Richland, WA 99352, USA}

\author{Sotiris Xantheas}
\affiliation{Pacific Northwest National Laboratory, Richland, WA 99352, USA}

\author{\"Ors Legeza}
\affiliation{Strongly Correlated Systems ``Lend\"{u}let'' Research group, Wigner Research Centre for Physics, H-1525, Budapest, Hungary}

\author{Libor Veis}
\email{libor.veis@jh-inst.cas.cz}
\affiliation{J. Heyrovsk\'{y} Institute of Physical Chemistry, Academy of Sciences of the Czech \mbox{Republic, v.v.i.}, Dolej\v{s}kova 3, 18223 Prague 8, Czech Republic}

\keywords{quantum chemistry, strong correlation, DMRG, parallelization, MPI, supercomputers}

\newcommand*{\tran}{^{\mkern-1.5mu\mathsf{T}}}

\begin{document}

\begin{abstract}
We present, to the best of our knowlegde, the first attempt to exploit the supercomputer platform for quantum chemical density matrix renormalization group (QC-DMRG) calculations. We have developed the parallel scheme based on the in-house MPI global memory library, which combines operator and symmetry sector parallelisms, and tested its performance on three different molecules, all typical candidates for QC-DMRG calculations. In case of the largest calculation, which is the nitrogenase FeMo cofactor cluster with the active space comprising 113 electrons in 76 orbitals and bond dimension equal to 6000, our parallel approach scales up to approximately 2000 CPU cores.
\end{abstract}

\maketitle

\section{Introduction}
\label{section_introduction}

The density matrix renormalization group (DMRG) method represents a very powerful approach originally developed for treatment of one-dimensional systems in solid state physics \cite{White-1992b, White-1993}.

Further success of DMRG in physics motivated its application also in quantum chemistry (QC) \cite{white_1999, Chan-2002a, Chan-2003, Legeza-2003a, Legeza-2003c, Legeza-2003b}, where it has shortly developed into an advanced multireference approach capable of going well beyond the limits of standard quantum chemical methods in problems where large complete active spaces (CAS) are mandatory and even reach the full configuration interaction (FCI) limit \cite{Legeza-2008, Marti-2010c, chan_review, wouters_review, yanai_review, Szalay-2015a}.

It has been applied on various problems ranging from
very accurate computations on small molecules \cite{Chan-2003, Chan-2004b, Sharma-2014},
extended \mbox{(pseudo-)} linear systems like polyenes, polyacenes, or graphene nanoribbons \cite{Hachmann-2007, Ghosh-2008, Mizukami-2013, Barcza-2013, Hu2015, Timar2016, amaya_2015},
transition-metal compounds \cite{Kurashige-2009, Marti-2010c, barcza_2011, boguslawski_2012, wouters_2014, amaya_2015, Nachtigallova2018},
or molecules containing heavy-element atoms which require relativistic four component treatment \cite{Knecht-2014, Battaglia2018}. Recently, the limits of the QC-DMRG method have been pushed by large scale computations of challenging bio-inorganic systems \cite{Kurashige-2013, sharma_2014b, nike, chan_new}.
During the past few years, several post-DMRG methods capturing the missing dynamic electron correlation on top of the DMRG wave function have also been developed \cite{Kurashige-2011, Saitow-2013, neuscamman_2010_irpc, sharma_2014c, Veis2016, Freitag2017}.

Regarding the parallelization strategies for QC-DMRG, the algorithms for shared \cite{Hager-2004}, as well as distributed \cite{Chan-2004, Kurashige-2009, Chan2016} memory architectures have been developed. The Chan's distributed approach \cite{Chan-2004} is based on parallelization over different terms in the Hamiltonian and assigns certain orbital indices (and the corresponding renormalized operators) to individual processors. An alternative approach of Kurashige \textit{et al.} \cite{Kurashige-2009} is based on parallelization over different symmetry sectors. Recently, the matrix-product-operator (MPO) inspired parallelization scheme employing the sum of operators formulation, which should result in lower inter-node communication requirements, was proposed \cite{Chan2016}.

Completely different approach than those presented so far was suggested by Stoudenmire and White \cite{Stoudenmire-2013}. This scheme relies on the observation that DMRG approximately preserves the reduced density matrix over regions where it does not sweep. In this approach, the lattice of orbitals is divided into several parts and sweeping on these parts is realized in parallel.

Extension of parallelization scheme of Ref.~\cite{Hager-2004} to a smart hybrid CPU-GPU implementation has also been presented, exploiting the power of both
CPU and GPU tolerating problems exceeding the GPU memory size~\cite{Nemes-2014}.
In DMRG, the iteartive construction of the Hamiltonian is decomposed into several independent matrix operations and each of these are further decomposed
into smaller independent tasks based on symmetries, thus diagonalization has been expressed as a single list of dense matrix operations.

There exist a few great QC-DMRG codes \cite{wouters_review} with different functionalities and most of them are open-source and available online. However, according to the best of our knowledge, none of them is truly massively parallel, i.e. can be run advantageously on hundreds or more than a thousand of CPU cores.
This article is thus a first attempt to port the QC-DMRG method to a supercomputer platform. Our parallel approach is similarly to the shared memory algorithm \cite{Hager-2004, Nemes-2014} based on merging of the operator and symmetry sector loops and employ the global memory model. It relies on a fast inter-node connection. 

The new C++ QC-DMRG implementation named \textsf{MOLMPS}\footnote{The \textsf{MOLMPS} code with all its functionalities will be presented in a different publication.} was created based on this parallel approach.

The paper is organized as follows: in section \ref{subsection_dmrg}, we give a brief overview of the QC-DMRG method. Since the \textsf{MOLMPS} program employes the renormalized operators rather than MPOs\footnote{It is just a matter of taste, both formulations are equivalent in terms of efficiency \cite{Chan2016}.}, the presentation is in the original renormalization group picture \cite{schollwock_2005}. Section \ref{subsection_parallel} contains the details of our parallel scheme and the computational details of our numerical tests are presented in section \ref{section_computational_details}. Section \ref{section_results}  summarizes the results with discussion, and  section \ref{section_conclusions} closes with conclusions and outlook.

\section{Theory}
\label{section_theory}

\subsection{QC-DMRG overview}
\label{subsection_dmrg}

In non-relativistic electronic structure calculations, one is interested in eigenvalues and eigenvectors of the electronic Hamiltonian with the following second-quantized structure \cite{ostlund_szabo}

\begin{eqnarray}
  \label{H2q} 
  H_{\text{el.}} & = & \sum_{\underset{\sigma \in \{ \uparrow, \downarrow\}}{pq=1}}^n h_{pq} a_{p \sigma}^\dagger a_{q \sigma} +  \nonumber \\ 
  & + & \sum_{\underset{\sigma, \sigma^{\prime} \in \{ \uparrow, \downarrow\}}{pqrs=1}}^n v_{pqrs} a_{p \sigma}^\dagger a_{q \sigma^{\prime}}^\dagger a_{r \sigma^{\prime}} a_{s \sigma} ,
\end{eqnarray}

\noindent
where $h_{pq}$ and $v_{pqrs}$ represent one and two-electron integrals in a molecular orbital (MO) basis,  which is for simplicity assumed to be restricted (e.g. restricted Hartree-Fock), $\sigma$ and $\sigma^{\prime}$ denote spin variables, and $n$ is the size of the MO space in which the Hamiltonian is diagonalized.

\begin{figure}[!ht]
  \hskip -3.5cm
  \begin{minipage}{0.5\textwidth}
    \includegraphics[width=12cm]{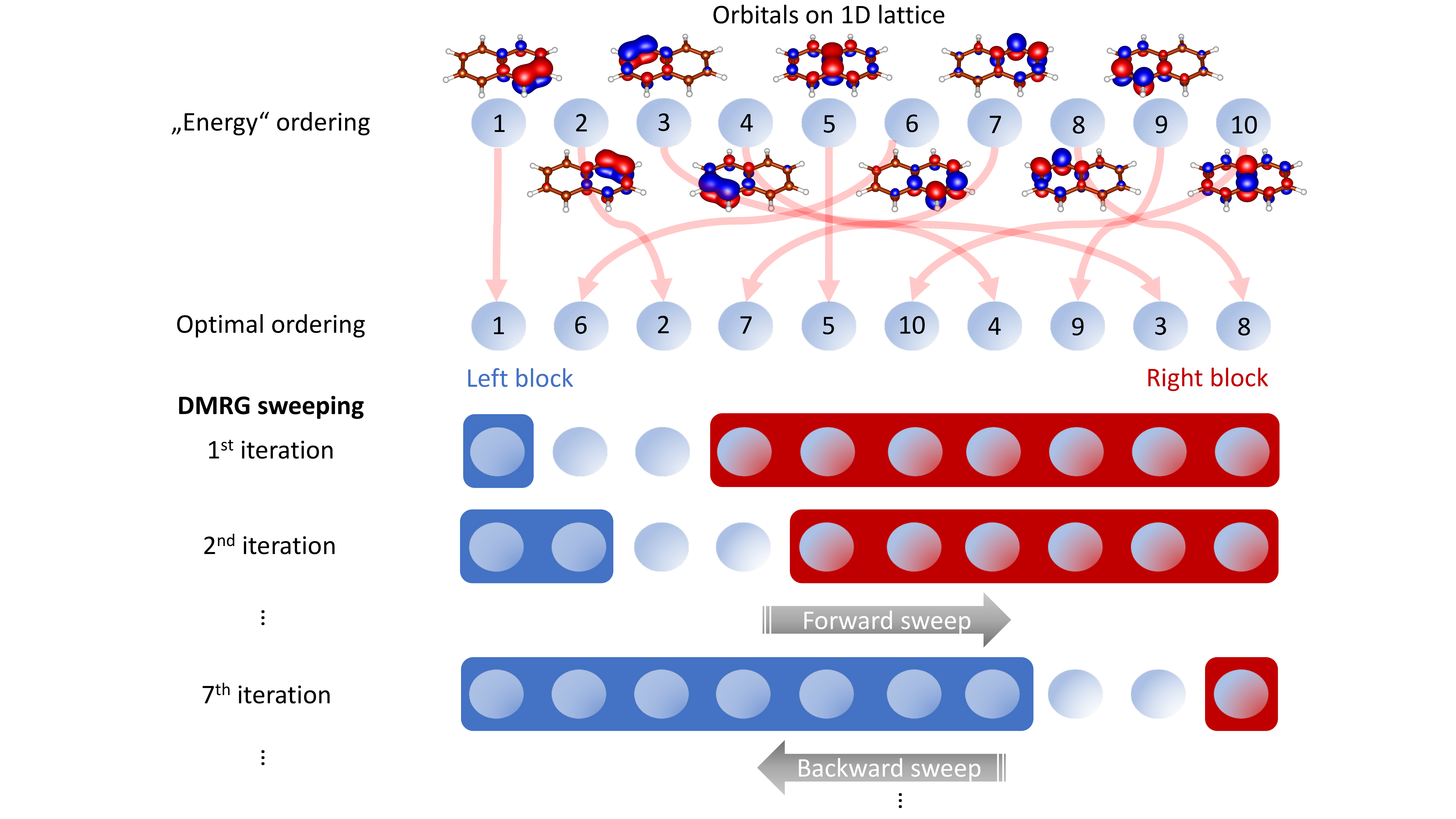}
  \end{minipage}
  \caption{The scheme of the DMRG sweep algorithm.}
  \label{dmrg_blocks}
\end{figure}

The very first step of a QC-DMRG calculation is to order the individual MOs on a one-dimensional lattice, putting mutually strongly correlated orbitals as close as possible, which may be carried out e.g. with the help of techniques developed in the field of quantum information \cite{Szalay-2015a, Legeza-2003b, rissler_2006, barcza_2011}. Then in the course of the practical two-site QC-DMRG sweep algorithm \cite{schollwock_2005}, $H_{\text{el.}}$ is being diagonalized in a vector space which is formed as a tensor product of the four spaces; so called left block, left site, right site, and right block. The sweep algorithm starts with just a single orbital in the left block, which is then enlarged in each DMRG iteration by one orbital up to the point, where the right block contains only a single orbital. The right block is being enlarged afterwards, see Figure \ref{dmrg_blocks}, and the sweeping is repeated until the energy is converged. There is in fact an analogy between the DMRG sweep algorithm and the Hartree-Fock self-consistent iterative procedure \cite{Chan-2008b}.

A single MO (site) may be empty, occupied by one $\alpha$ or $\beta$ electron, or doubly occupied. The corresponding vector space is thus spanned by the four basis states $\{ \ket{0}, \ket{\uparrow}, \ket{\downarrow}, \ket{\uparrow \downarrow}\}$ and is complete. The matrix representations of creation operators in this basis read

\begin{equation}
  \label{creation_ops}
  a^{\dagger}_{\uparrow} = \begin{pmatrix} 0 & 0 & 0 & 0 \\ 0 & 0 & 0 & 0 \\ 1 & 0 & 0 & 0 \\ 0 & 1 & 0 & 0 \end{pmatrix}, \quad a^{\dagger}_{\downarrow} = \begin{pmatrix} 0 & 0 & 0 & 0 \\ 1 & 0 & 0 & 0 \\ 0 & 0 & 0 & 0 \\ 0 & 0 & -1 & 0 \end{pmatrix} .
\end{equation}

Enlarging the left or right block with $M$ basis states by one MO as mentioned above, would without any truncation lead to the new $4M$-dimensional vector space and when repeated to the curse of dimensionality. The essence of the DMRG algorithm \cite{White-1992b, White-1993} is indeed to determine the optimal left and right block many-electron basis with bounded dimension $M$, so called bond dimension \cite{White-1992b, White-1993, schollwock_2005}. When forming e.g. the left enlarged block containing $p$ orbitals, the full vector space is spanned by $\{l_{p-1}\} \otimes \{s\}$, where $\{l_{p-1}\}$ denotes the basis of the left block with $p-1$ orbitals and $\{s\}$ the basis of the added ($p$-th) orbital (site). In order to keep the dimension $M$, the new basis must be truncated in a following way

\begin{equation}
  \label{renorm}
  \ket{l_p} = \sum_{l_{p-1} s} O_{l_{p-1} s, l_p}^L \ket{l_{p-1}} \otimes \ket{s},
\end{equation}

\noindent
where $\mathbf{O}^L$ is the $4M \times M$ left block renormalization matrix. 

In case of the DMRG algorithm, the determinant representation of the complicated many-electron basis is not stored, instead the matrix representations of second-quantized operators needed for the action of the Hamiltonian (\ref{H2q}) on a wave function are formed and stored. For a single orbital, all the required operator matrices can be formed from matrices in (\ref{creation_ops}) by matrix transpositions, multiplications with appropriate MO integrals, and matrix-matrix multiplications. 

For the block of orbitals, the situation is more complicated. Since the renormalized many-electron basis is not complete, one cannot store only matrices of creation (or annihilation) operators acting on individual orbitals of the given block and form matrices of operators corresponding to the strings of second-quantized operators appearing in (\ref{H2q}) by their multiplications. In fact, one has to form all operator intermediates necessary for the action of the Hamiltonian (\ref{H2q}) on a wave function.

Projecting the Schr\"{o}dinger equation onto the product space of the left block, left site, right site, and right block $\big(\{l\} \otimes \{s_l\} \otimes \{s_r\} \otimes \{r\}\big)$, we have the effective equation

\begin{equation}
  \label{h_mat}
  \mathbf{H}_{\text{el.}} \bm{\psi} = E \bm{\psi},
\end{equation}

\noindent
where $\bm{\psi}$ are the expansion coefficients of the wave function, thus

\begin{equation}
  \label{wf}
  \ket{\Psi} = \sum_{l s_l s_r r} \psi_{l s_1 s_2 r} \ket{l} \otimes \ket{s_l} \otimes \ket{s_r} \otimes \ket{r}.
\end{equation}

In order to reduce the number of matrix-matrix multiplications during the action of the Hamiltonian on a wave function, which are the most CPU-demanding tasks, the efficient QC-DMRG codes work with the so called pre-summed (or partially summed) operators \cite{Xiang-1996}, i.e. intermediates formed by contraction of operator matrices with MO integrals. For example in the left block

\begin{equation}
  \mathcal{A}^{\uparrow \uparrow}_{rs} = \sum_{pq \in \text{left}} v_{pqrs} a^{\dagger}_{p\uparrow} a^{\dagger}_{q\uparrow}, \qquad rs \not \in \text{left}.
\end{equation}

\noindent
$\mathcal{A}^{\uparrow \uparrow}_{rs}$ are examples of the left block two-index pre-summed operators which together with $a_{r\uparrow} a_{s\uparrow}$ acting on the two sites or the right block (plus Hermitian conjugate terms) contribute to the $(\uparrow \uparrow \uparrow \uparrow)$-part of the two-electron Hamiltonian interaction term

\begin{eqnarray}
  H^{\uparrow \uparrow \uparrow \uparrow}_{\text{int}} & \ni & \sum_{\substack{pq \in \text{left} \\ rs \not \in \text{left}}} v_{pqrs} a^{\dagger}_{p\uparrow} a^{\dagger}_{q\uparrow}  a_{r\uparrow} a_{s\uparrow} = \nonumber \\
   & = & \sum_{rs \not \in \text{left}} \mathcal{A}^{\uparrow \uparrow}_{rs} a_{r\uparrow} a_{s\uparrow} .
\end{eqnarray}

\noindent
Notice that the four-index summation have been replaced by the two-index one\footnote{Also the two-index pre-summed operators are formed in such a way, that the contraction with MO integrals is performed for the larger block, keeping the remaining two sums as short as possible \cite{Kurashige-2009, Chan2016}.}. When employing the partial summations, all the operators that build up the Hamiltonian are at most two-index (normal or pre-summed) \cite{white_1999}.

The full Hamiltonian matrix (\ref{h_mat}) is not formed, instead the tensor product structure of the vector space is employed. For example let us assume that we have the above mentioned contributing term, where $\mathcal{A}^{\uparrow \uparrow}_{rs}$ in the left block is accompanied by $a_{r\uparrow} a_{s\uparrow}$ in the right one;
$
\big(\mathcal{A}^{\uparrow \uparrow}_{rs} \otimes I \otimes I \otimes a_{r\uparrow} a_{s\uparrow}\big),
$
$I$ being the identity

\begin{widetext}
  \begin{equation}
    \bra{l^{\prime}} \otimes \bra{s_l^{\prime}} \otimes \bra{s_r^{\prime}} \otimes \bra{r^{\prime}} \big(\mathcal{A}^{\uparrow \uparrow}_{rs} \otimes I \otimes I \otimes a_{r\uparrow} a_{s\uparrow}\big) \ket{l} \otimes \ket{s_l} \otimes \ket{s_r} \otimes \ket{r} = \bra{l^{\prime}} \mathcal{A}^{\uparrow \uparrow}_{rs} \ket{l} \bra{r^{\prime}} a_{r\uparrow} a_{s\uparrow} \ket{r} \delta_{s_l^{\prime} s_l} \delta_{s_r^{\prime} s_r}.
  \end{equation}
\end{widetext}

\noindent
The action of this Hamiltonian term on a trial wave function vector ($\phi_{l s_1 s_2 r}$) needed for the iterative diagonalization solvers like Davidson algorithm \cite{Davidson-1975}, can be therefore compiled only from the knowledge of composing operator matrices in the basis of the individual blocks (left, right, or sites).

To complete the overview of the QC-DMRG algorithm, it remains to define the renormalization matrix (\ref{renorm}).
As argued by White \cite{White-1992b, White-1993}, it is optimal to make the renormalization matrix $\mathbf{O}^L$ (or $\mathbf{O}^R$) from the $M$ eigenvectors of the left (or right) enlarged block reduced density matrix with the largest eigenvalues. When the wave function expansion coefficients $\psi_{l s_l s_r r}$ (\ref{wf}) are reshaped into the matrix form $\psi_{(l s_l), (s_r r)}$, the aforementioned reduced density matrices can be computed in the following way

\begin{eqnarray}
  \bm{\rho}^L & = & \bm{\psi} \bm{\psi}^{\dagger}, \\
  \bm{\rho}^R & = & \bm{\psi}^{\dagger} \bm{\psi}.
\end{eqnarray}

\noindent
For the transition to the next iteration, all operator matrices formed for the enlarged block, e.g. in $\ket{l_{p-1}} \otimes \ket{s}$ basis for the forward sweep (\ref{renorm}), have to be renormalized

\begin{equation}
  \label{renorm_mat}
  \bm{A}^{\prime} = (\bm{O}^L)^\dagger \bm{A} \bm{O}^L ,
\end{equation}

\noindent
where $\bm{A}$ represents an operator matrix in the non-truncated ($4M$-dimensional) basis and $\bm{A}^{\prime}$ is the renormalized matrix representation in the truncated ($M$-dimensional) basis.

Another ingredient of the efficient QC-DMRG code is a proper handling of quantum symmetries \cite{McCulloch-2000,Toth-2008}. Currently, the \textsf{MOLMPS} code employes $U(1)$ symmetry\footnote{Point group symmetry may be employed as well, it is however useful only in case of small symmetric molecules. Otherwise, the localized or split-localized MO basis, which break the point group symmetry, are typical in QC-DMRG calculations \cite{amaya_2015}. Localized or split-localized MO basis were also used in the presented numerical examples.}, however the $SU(2)$ (spin-adapted version) \cite{Sharma-2012a, Wouters-2014a, keller_2016} is under development and it will not affect the parallel scheme presented in the next subsection.

We employ $U(1)$ symmetries to restrict the total number of $\alpha$ and $\beta$ electrons (or equivalently spin projection $M_S$). As usually, all left and right block basis states, as well as the site basis states, are grouped into symmetry sectors sharing the number of $\alpha$ ($n_{\uparrow}$) and $\beta$ ($n_{\downarrow}$) electrons. Only the non-zero blocks of operator matrices are stored in the form of dense matrices together with the necessary information about the symmetry sectors.

When expanding the wave function (\ref{wf}) in the symmetry-sector decomposed form of the left block, left site, right site, and right block basis, only those sectors whose $n_{\uparrow}$  and $n_{\downarrow}$ sum up to the correct total numbers ($n_{\uparrow}^{\text{tot.}}, n_{\downarrow}^{\text{tot.}}$) contribute. Moreover in case of the non-relativistic QC-DMRG method, for which $n_{\uparrow}$  and $n_{\downarrow}$ are good quantum numbers\footnote{This is not the case of the four-component relativistic QC-DMRG, where only the total number of electrons is a good quantum number \cite{Knecht-2014,Battaglia2018}.}, all the symmetry sectors of a single site are one-dimensional \cite{Szalay-2015a} and we can write

\begin{equation}
  \label{wf_sec}
  \ket{\Psi} = \sideset{}{'}\sum_{abcd} \sum_{\substack{l \in a \\ r \in d}} \psi_{l r}^{abcd} \ket{l} \otimes \ket{s_l^b} \otimes \ket{s_r^c} \otimes \ket{r},
\end{equation}

\noindent
where $a,b,c,d$ denote indices of the left block, left site, right site, and right block symmetry sectors and primed summation symbol stands for the restricted summation for which holds

\begin{eqnarray}
  n_{\uparrow} (a) + n_{\uparrow} (b) + n_{\uparrow} (c) + n_{\uparrow} (d) = n_{\uparrow}^{\text{tot.}} , \\
  n_{\downarrow} (a) + n_{\downarrow} (b) + n_{\downarrow} (c) + n_{\downarrow} (d) = n_{\downarrow}^{\text{tot.}} .
\end{eqnarray}

\noindent
$\ket{s_l^b}$ and $\ket{s_r^c}$ in (\ref{wf_sec}) denote $b^{\text{th}}$ symmetry sector left site basis state and $c^{\text{th}}$ symmetry sector right site basis state.

In summary, one QC-DMRG iteration is composed of the three main steps, whose parallelization is discussed in detail in the next subsection, namely: (a) formation of pre-summed operators, (b) Hamiltonian diagonalization, and (c) operators renormalization. The overall cost of the QC-DMRG computation is $\mathcal{O}(M^2n^4) + \mathcal{O}(M^3n^3)$, where the first term corresponds to the formation of pre-summed operators whereas the second one to the Hamiltonian diagonalization and renormalization \cite{white_1999}.
  
Last but not least, we would like to briefly mention the connection to the matrix product state (MPS) wave function form \cite{Schollwock-2011}. In fact, the MPS matrices are nothing, but reshaped renormalization matrices (\ref{renorm}), and they are easily obtainable from the DMRG sweep algorithm \cite{Chan2016}. They can be used e.g. for efficient calculations of correlation functions or a subset of the FCI expansion coefficients, which may be employed for the purposes of the tailored coupled cluster methods \cite{Veis2016, Veis2016corr, Veis2018, Faulstich2019, Antalik2019, porph}. If we reshape $O_{l_{p-1} s, l_p}^L$ from (\ref{renorm}) to $L^s_{l_{p-1}, l_p}$ (i.e. $4M \times M$ matrix into $M \times M \times 4$ tensor) and similarly for the right block matrix, then starting with the wave function expansion in the renormalized basis (\ref{wf}) and recursive application of (\ref{renorm}) would lead to

\begin{eqnarray}
  \ket{\Psi_{\text{MPS}}} & = & \sum_{\{ s \}} \bm{L}^{s_1} \bm{L}^{s_2} \ldots \bm{\psi}^{s_i s_{i+1}} \ldots \bm{R}^{s_n} \cdot \nonumber \\
   & \cdot & \ket{s_1 s_2 \ldots s_n}
\end{eqnarray}

\noindent
which is the two-site MPS form of the DMRG wave function.

\subsection{Parallel scheme}
\label{subsection_parallel}

\begin{figure}[!ht]
  \hskip -1cm
  \begin{minipage}{0.5\textwidth}
    \includegraphics[width=10cm]{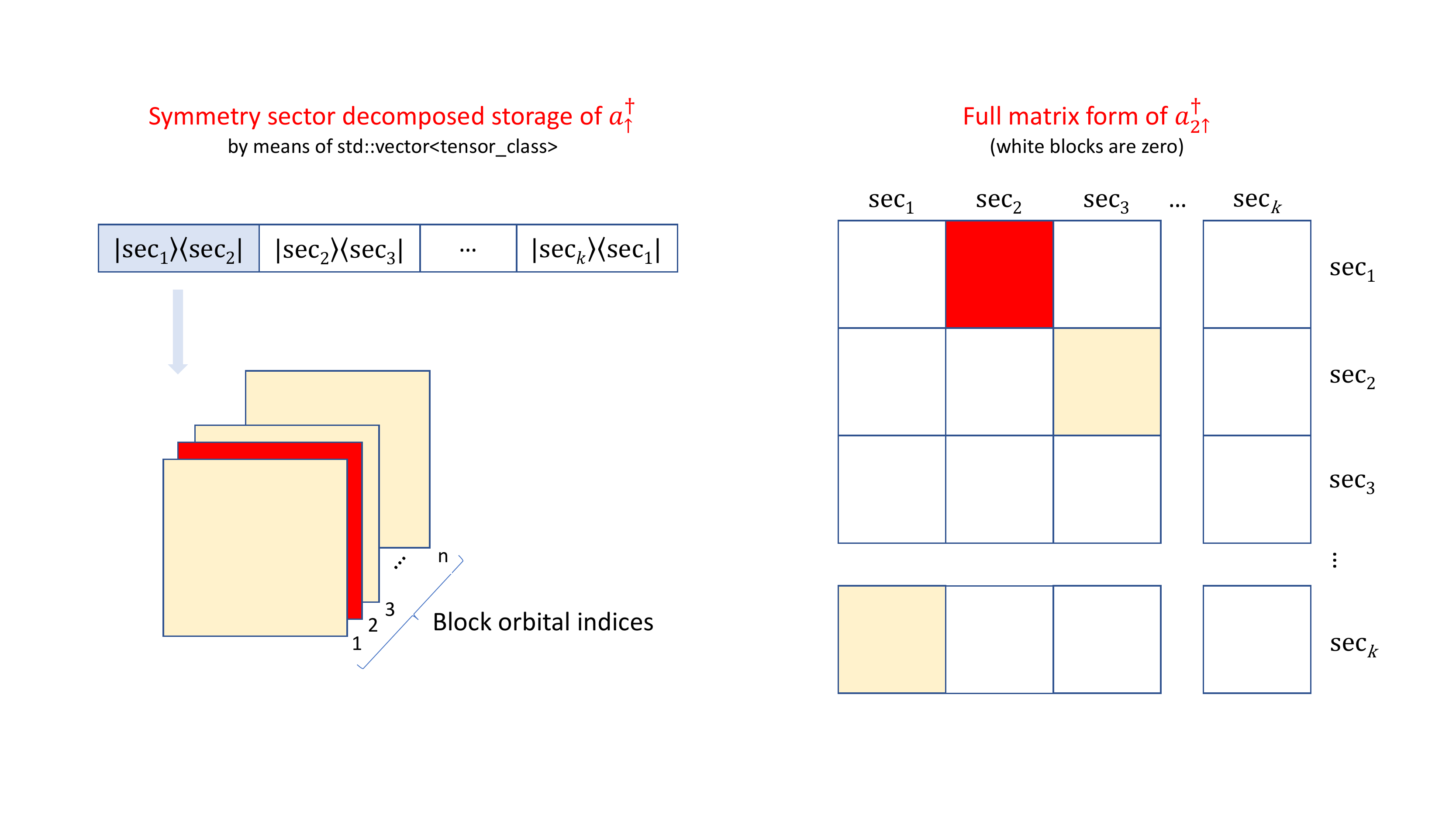}
  \end{minipage}
  \caption{The operator class storage demonstrated on the example of the block operator $a^{\dagger}_{\uparrow}$.}
  \label{opclass}
\end{figure}

Before discussing the parallel scheme, let us first briefly describe, how the data is stored in \textsf{MOLMPS}, in particular the operators. As is usual, we employ the sparsity generated by quantum symmetries mentioned in the previous subsection. The operator class contains information about individual symmetry sectors and the corresponding dense matrices. 
For these dense entities, 
we have developed our own lightweight tensor library
which can work with up to three-legged dense tensors and serves as a wrapper to BLAS and LAPACK. In case of the block operators with one or two orbital indices (normal or pre-summed), the third index of the dense tensor corresponds to the orbital index / pair of indices. The dense tensors of individual symmetry sectors are stored in the vector container of the C++ standard library as is depicted in Figure \ref{opclass}.

Our parallel approach is based on our own MPI global memory (GM) library.
It relies on a fast inter-node connection (e.g. Infiniband, Omni-Path, Fibre Chanel) which is common for all modern supercomputer architectures.
The data distribution and handling are managed by a GM class.
The GM class instance envelops a group of tensors, which are distributed under the same conditions, e.g. the operators of a given (left or right) block and carries all the information about the distribution.

In order to minimize the amount of data stored in the memory, we employed the MPI shared memory (SHM) model (introduced with MPI version 3), so only one copy of each tensor/matrix is stored per node. All processes
on a given node can access these data directly without using a remote access. Remote processes can access these data using inter-node communicators by RMA calls.

For distribution, the GM supports two models.
The first one, which is suitable for the smallest arrays, is the local data model, where selected tensors are available locally on all nodes.
This significantly reduces communication for the reasonable price of a slightly higher memory requirements.
User can in fact specify a threshold for the array size, below which this model is employed.
This model is by default used for example for the Krylov vectors during the Davidson diagonalization \cite{Davidson-1975}.

The second option is the global model. It is suitable for large arrays and the data are evenly distributed among nodes (in the MPI SHM regime).
The distribution is performed over all available nodes in such a way that a balanced load on nodes is ensured.
This model is typically used for dense tensors of the individual sectors of the left and right block operators in case of larger calculations (active space sizes and bond dimensions).
When the tensors fit into the memory of a single node, the above mentioned local memory approach is certainly more advantageous (see the results section).
Indeed, any intermediate of the DMRG calculation may be treated in a different data model, purely based on the amount of free computer memory, just to maximize data locality.

The already mentioned in-house dense tensor library provides a templated tensor class for simple tensor handling in combination with the GM class. The tensor class consists
of a pointer to the data array and descriptors, which involve dimensions and also a pointer to the GM class. While descriptors are available locally for each process, data arrays are handled by GM class.
This design of the code substantially simplifies code design and data handling.

When performing any operation with dense tensors, \texttt{get\_data()} call recognizes whether the data are stored locally or remotely and then return data pointer or fetch the data from remote location first.
We also introduced pre-fetching of tensors (or parts of it) needed for a calculation where the same sectors are involved multiple times.
\texttt{fetch\_data()} or \texttt{fetch\_slices()} fetch the data from remote location and assign it temporarily to a corresponding
tensor class. When the data are no longer needed, the allocated memory is freed.

Regarding the sources of parallelism, we combine operator and symmetry sector parallelisms similarly to the simpler shared memory approach \cite{Hager-2004, Nemes-2014}, in order to generate a large-enough number of tasks (dense matrix-matrix operations), which can be executed in parallel.
All three main steps are task-based parallelized.

\subsubsection{Hamiltonian diagonalization}

In case of the iterative Hamiltonian diagonalization (\ref{h_mat}) by means of the Davidson \cite{Davidson-1975} or similar algorithms, the Hamiltonian is applied sequentionally on a trial wave function vector. This action is composed of a large number of operator combinations

\begin{eqnarray}
  \ket{\widetilde{\Phi}} & = & H_{\text{el.}} \ket{\Phi} \\
   & = & \sum_{\alpha} \big( A^{(\alpha)}_{\text{l}} \otimes A^{(\alpha)}_{\text{s}_{\text{l}}} \otimes A^{(\alpha)}_{\text{s}_\text{r}} \otimes A^{(\alpha)}_{\text{r}} \big) \ket{\Phi} , \nonumber
\end{eqnarray}

\noindent
where $\alpha$ denotes a given operator combination and 
$A_\text{l}$, $A_{\text{s}_\text{l}}$, $A_{\text{s}_\text{r}}$, and $A_\text{r}$ 
represent the left block, left site, right site, and right block operators, respectively.
One such operator combination is e.g. the term
$
\mathcal{A}^{\uparrow \uparrow}_{rs} \otimes I \otimes I \otimes a_{r\uparrow} a_{s\uparrow},
$
which was mentioned earlier.

When taking into account the symmetry sectors (\ref{wf_sec}), then it holds for the expansion coefficients of the resulting vector $\ket{\widetilde{\Phi}}$ 

\begin{eqnarray}
  \widetilde{\phi}_{l^{\prime}r^{\prime}}^{a^{\prime}b^{\prime}c^{\prime}d^{\prime}} & = & \sum_{\alpha} \sideset{}{'}\sum_{abcd} \sum_{\substack{l \in a \\ r \in d}} \big[ A^{(\alpha)}_{\text{l}} \big]_{l^{\prime} l}^{a\rightarrow a^{\prime}} \big[ A^{(\alpha)}_{\text{s}_\text{l}} \big]^{b\rightarrow b^{\prime}} \cdot \nonumber \\
  & \cdot &  \big[ A^{(\alpha)}_{\text{s}_\text{r}} \big]^{c\rightarrow c^{\prime}} \big[ A^{(\alpha)}_{\text{r}} \big]_{r^{\prime} r}^{d\rightarrow d^{\prime}} \phi^{abcd}_{lr}, \nonumber \\[0.5cm]
  & & l^{\prime} \in a^{\prime} \text{ and } r^{\prime} \in d^{\prime} .
\end{eqnarray}

\noindent
The superscript of type $a \rightarrow a^{\prime}$ labels the symmetry sector of a given operator to which the operator matrix elements belong. If we formally gather all symmetry sector indices to $s$ (and $s^{\prime}$), merge elements of the site operators (scalars) into a multiplication factor $f(s)$, and reshape the wave function expansion coefficients into a matrix form, we can write

\begin{equation}
  \label{dgemm}
  \widetilde{\bm{\phi}}^{s^{\prime}} = \sum_{\alpha s} f(s) \cdot \bm{A}_{\text{l}}^{(\alpha, s, s^{\prime})} \cdot \bm{\phi}^s \cdot \big( \bm{A}_{\text{r}}^{(\alpha, s, s^{\prime})} \big) \tran .
\end{equation}

\noindent
The action of the Hamiltonian on a trial wave function vector thus comprise a huge number of dense matrix-matrix multiplications.

In our approach, before the Davidson algorithm is started, a huge task list which combines the loops over operator combinations ($\alpha$) and symmetry sectors ($s, s^{\prime}$) is generated. Since different terms from this task list may write to the same wave function output sector ($s^{\prime}$), each MPI process has its own copy of $\bm{\widetilde{\phi}}$ and we use the reduce function. 

In case of the GM model, the individual operator combinations acting on both (left and right) blocks at the same time do not involve orbital indices. So for example instead of the aforementioned term
$
\mathcal{A}^{\uparrow \uparrow}_{rs} \otimes I \otimes I \otimes a_{r\uparrow} a_{s\uparrow}
$,
we have 
$
\mathcal{A}^{\uparrow \uparrow} \otimes I \otimes I \otimes a_{\uparrow} a_{\uparrow}
$
and the loop over $rs$ is performed sequentionally during the task execution. It is organized this way to avoid fetching of small memory chunks and with the view of a further GPU acceleration in future [parallel execution of matrix-matrix multiplications (\ref{dgemm}) performed on different slices of the same dense tensors] \cite{Nemes-2014}.

In order to exploit data locality at the maximum, 
we have developed the semi-dynamic scheduler.
It considers where data for individual tasks are stored and also involves independent counters specific for each node. As the result, the group of tasks is assigned to a node, where their execution will cause the minimum amount of communication. The group of tasks are executed locally on a given node with a dynamical task distribution among local processes. In case of tasks involving tensors from both (left and right) blocks, the execution node is selected based on the storage of the larger from both tensors to minimize the amount of data being fetched.


\subsubsection{Operators renormalization}

Also in case of the renormalization, we generate a huge task list, which combines the loops over different operators to be renormalized and their symmetry sectors. These tasks are completely independent and can be executed in parallel. During a given task execution, 
the complete sector matrix of the newly formed operator corresponding to the non-truncated enlarged basis is formed according to the blocking tables\footnote{They contain the information about how the individual operators are combined during blocking, i.e. when the block is enlarged by a site.} and
renormalized by the sector matrices of the renormalization operator (\ref{renorm_mat}).

In case of the GM model, we again employ the semi-dynamic scheduler, which now considers the locality of the newly formed operator sector matrices. The block operator sector matrices needed for the blocking have to be fetched, which naturally requires much more communication than the Hamiltonian diagonalization \cite{Chan-2004}.
In order to decrease the communication, we do not work with individual slices of dense tensors of the newly formed operators (corresponding to different orbital indices), but rather group them into larger chunks. We also fetch the whole dense tensors of the block operators and order the tasks so that they may be re-used for subsequent tasks. It is at the cost of a slightly higher memory requirements.

In case of the local memory model, the situation is much simpler since there is no need for fetching of tensors of the block operators (everything is available on all nodes). The difference, however, is that we have to update all nodes with the newly formed sector matrices. Because in the local memory model, the dense tensors of the newly formed operators are stored in consecutive arrays, it can be done efficiently by means of the accumulate function.

\subsubsection{Operators pre-summation}

Since we fully employ the tensor product structure of the vector space of the two blocks and two sites, we form also additional pre-summed operators on-the-fly, during preparing the action of the Hamiltonian on a trial wave function. These operators are formed in order to minimize the number of matrix-matrix multiplications during the aforementioned Hamiltonian diagonalization and they are not renormalized and stored. 

As an example, let us consider the term

\begin{equation}
  \label{presum}
  \sum v_{pqrs} a^{\dagger}_{p\uparrow} \otimes a^{\dagger}_{q\uparrow} \otimes I \otimes a_{r\uparrow} a_{s\uparrow},
\end{equation}

\noindent
where $p$ belongs to the left block, $q$ to the left site, and $r$ with $s$ to the right block. If we form the following operators in the left block

\begin{equation}
  \mathcal{A}^{\text{tmp}}_{rs} = \sum_{\substack{p \in \text{left} \\ q = \text{left site} \\ rs \in \text{right}}} v_{pqrs} a^{\dagger}_{p\uparrow},
\end{equation}

\noindent
we can rewrite Eq. \ref{presum} as

\begin{equation}
  \sum_{rs \in \text{right}} \mathcal{A}_{rs}^{\text{tmp}} \otimes a^{\dagger}_{q\uparrow} \otimes I \otimes a_{r\uparrow} a_{s\uparrow}.
\end{equation}

In our approach, we first generate the list of all operators to be formed on-the-fly and then,
in line with previous subsections, generate a huge task list combining all the pre-summed operators and their symmetry sectors.

Not the local memory model, neither the GM model requires network communication. It is because in the GM model, we form the symmetry sector matrices of the pre-summed operators on the same nodes where the given symmetry sector tensors of the original operators, whose slices are multiplied by MO integrals and summed, are stored.

In case of the local memory model, all nodes have to be updated with the newly formed sector matrices, which is done in the same way as in case of the renormalization.

\section{Computational details}
\label{section_computational_details}

\begin{figure*}[!ht]
  \subfloat[Fe(II)-porphyrin model \label{FeP}]{%
    \includegraphics[width=4cm]{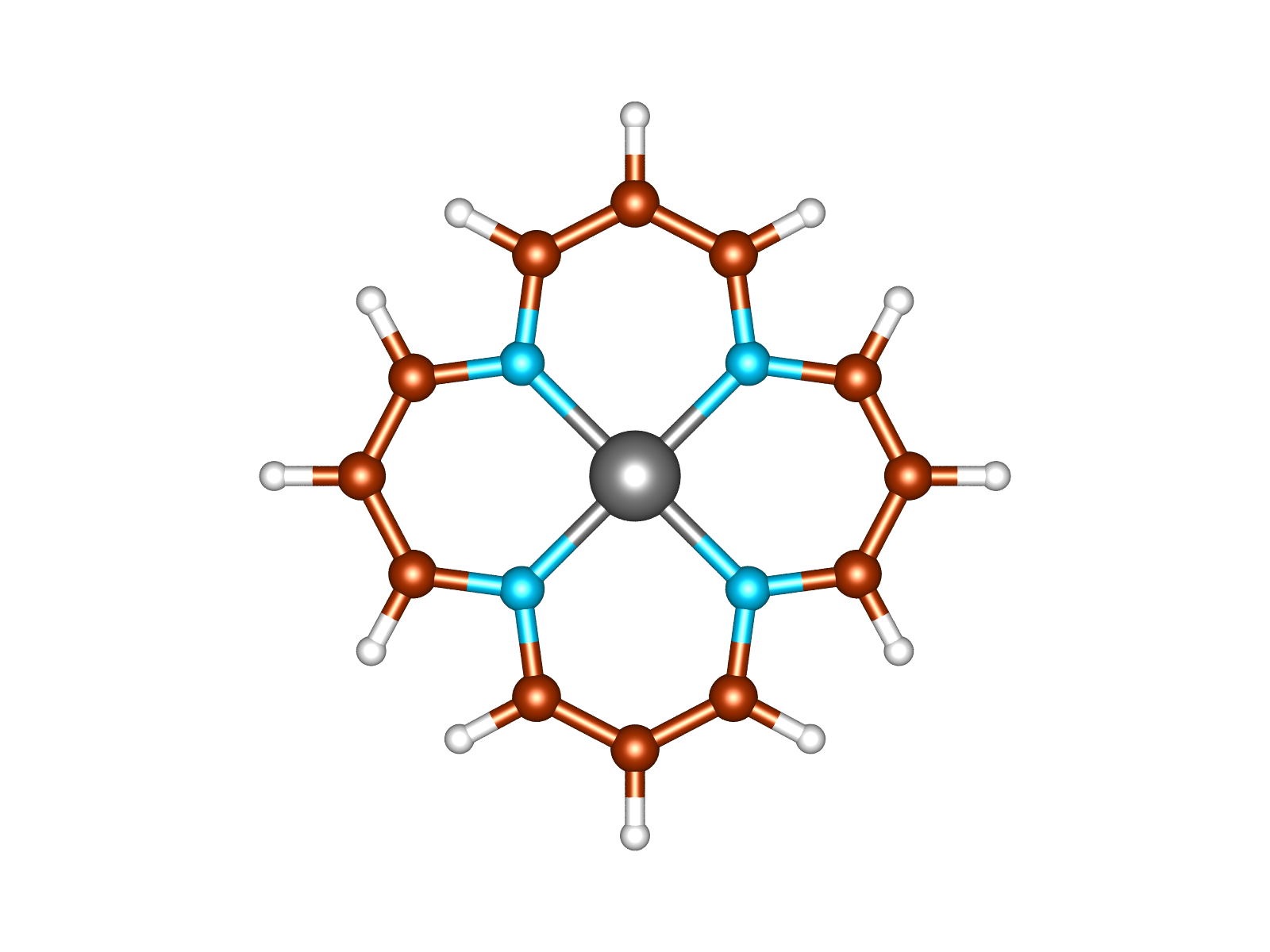}
  }
  \hfill
  \subfloat[Defected $\pi$-conjugated anthracene tetramer \label{a_56}]{%
    \includegraphics[width=8cm]{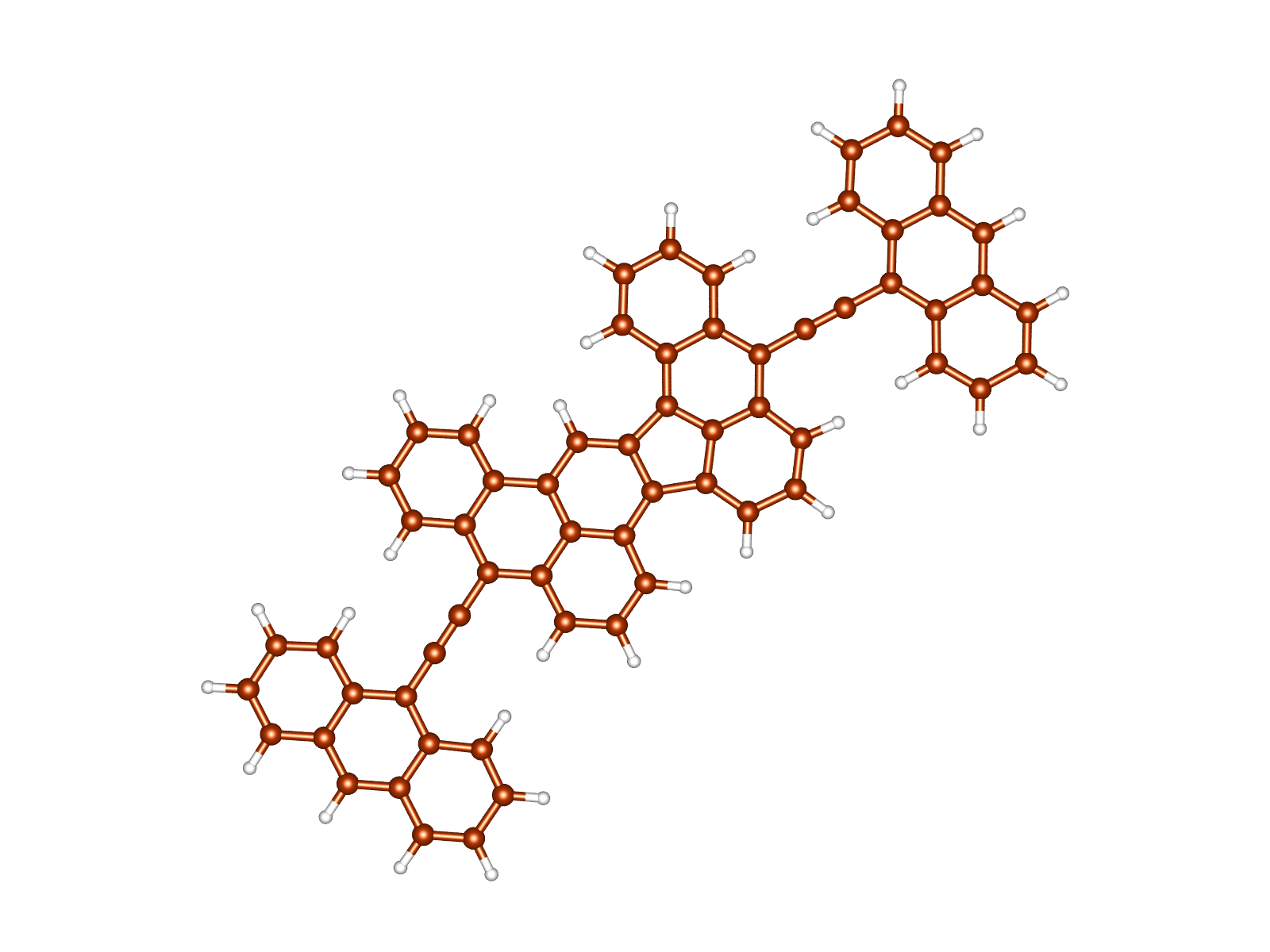}
  }
  \hfill
  \subfloat[FeMoco cluster \label{femoco}]{%
    \includegraphics[width=5cm]{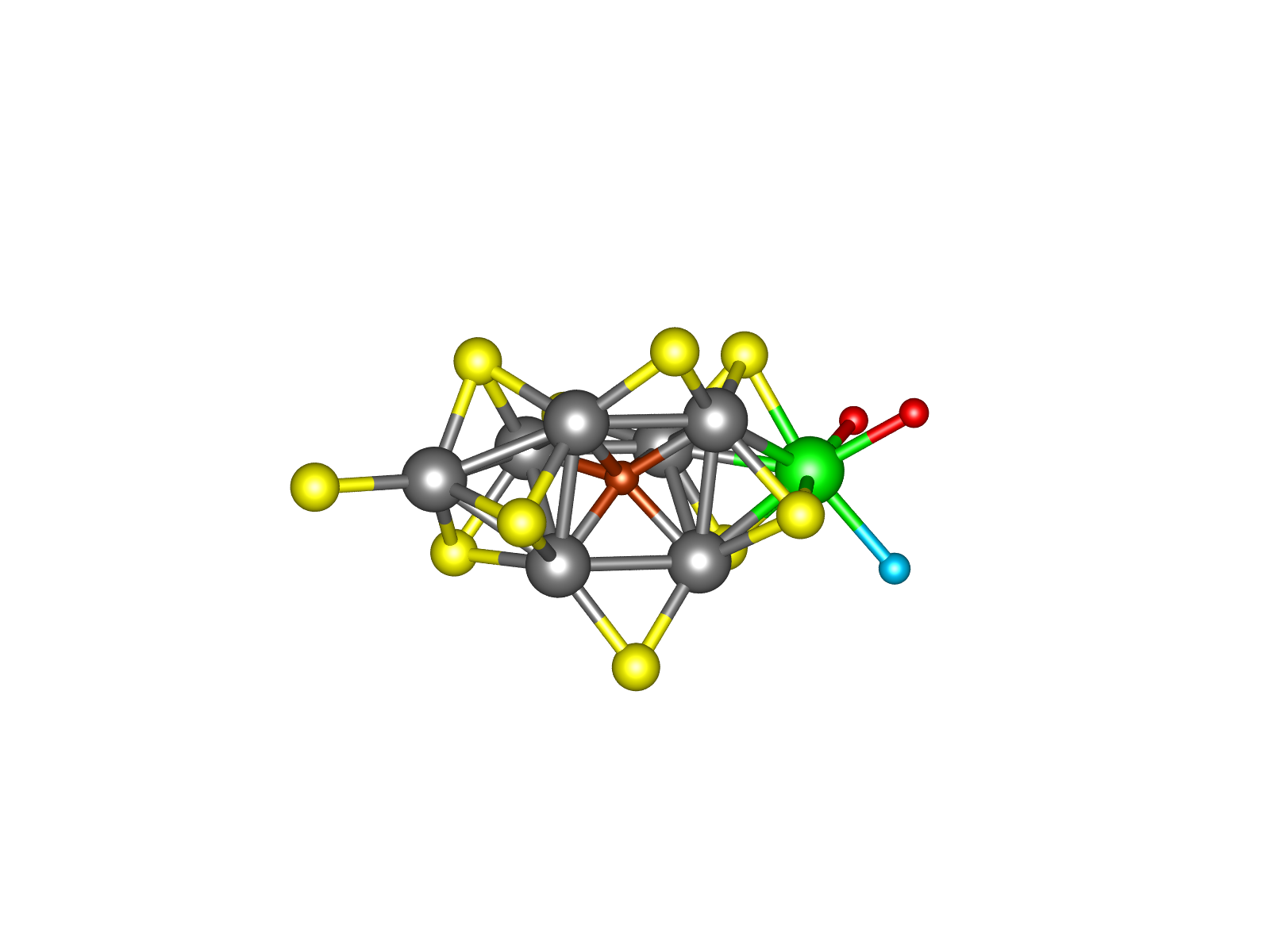}
  }
  \caption{Structures of the molecules for which the parallel QC-DMRG scaling has been studied. Notice that the FeMoco cluster is for clarity not complete, only the atoms of the ligands which are directly bonded to Fe and Mo atoms are displayed. Atom colors: nitrogen - blue, sulphur - yellow, oxygen - red, carbon - brown, hydrogen - white, iron - grey, molybdenum - green.}
\end{figure*}

We have tested the parallel scheme presented in the previous section on three different molecules, all typical candidates for the QC-DMRG computations, namely: Fe(II)-porphyrin model (Figure \ref{FeP}), extended $\pi$-conjugated system (Figure \ref{a_56}), and FeMoco cluster (Figure \ref{femoco}).

Fe(II)-porphyrin model was selected, because it was demonstrated by Li Manni \textit{et al.} \cite{LiManni2018, LiManni2019} that apart from $3d$, $4d$, and $4s$ orbitals of the Fe center and $\sigma(\text{Fe-N})$ orbital, inclusion of all $\pi$ orbitals from the porphyrin ring into the active space is necessary for a quantitative determination of its ground state, leading to CAS(32,34). In our recent DMRG-DLPNO-TCCSD(T) study \cite{porph}, we have optimized this CAS orbitals for the lowest triplet and quintet states and different geometries by means of the state specific DMRG-CASSCF method in TZVP basis. In the present study, we have tested the parallel QC-DMRG scaling on the above mentioned triplet state DMRG-CASSCF(32,34)/TZVP orbitals optimized at the geometry used in Li Manni \textit{et al.} works \cite{LiManni2018, LiManni2019}. The triplet state was chosen as it is more correlated than quintet \cite{LiManni2018, porph} and the active space orbitals were split-localized \cite{amaya_2015}.

The second system selected for the scaling tests is the defected $\pi$-conjugated anthracene tetramer (Figure \ref{a_56}). It is a representative of $\pi$-conjugated hydrocarbons (linear or quasi linear), a group of molecules frequently studied by means of QC-DMRG calculations in the C-atom $p_z$ active space \cite{Hachmann-2007, Ghosh-2008, Mizukami-2013, Hu2015, amaya_2015}. We have recently studied the ground state of the above mentioned defected anthracene tetramer with the QC-DMRG method since this and similar species often appear as unwanted by-products during  on-surface synthesis of ethynylene‐bridged anthracene polymers \cite{jelinek}. Depending on the defect, such species may exhibit peculiar electronic structure properties. For the present study, 
we have employed the UB3LYP optimized geometry and built the active space from ROHF/cc-PVDZ C-atom $p_z$ orbitals, which corresponds to CAS(63,63). The active space orbitals were fully localized \cite{amaya_2015}.

The last system is the nitrogenase FeMo cofactor (FeMoco) cluster (Figure \ref{femoco}, notice that for clarity reasons the structure is not complete, only the atoms of the ligands which are directly bonded to Fe and Mo atoms are displayed), which is undoubtedly one of the most challenging problems of the current computational chemistry. Its importance is proved by the fact that FeMoco is responsible for the nitrogen reduction during the process of nitrogen fixation under ambient conditions in certain types of bacteria \cite{femoco}. In contrast, the industrial Haber-Bosch process to produce ammonia (mainly for fertilizers) is very energetically demanding. In fact, the electronic structure of the FeMo cofactor remains poorly understood \cite{nike, chan_new}. Reiher \textit{et al.} proposed the model of FeMoco with the active space containing 54 electrons in 54 orbitals in the context of simulations on quantum computers \cite{reiher_pnas}. Recently, it was shown on a different model that the larger active space, in particular CAS(113, 76) is necessary for the correct open-shell nature of its ground state \cite{nike}. For our benchmark tests, we have employed the integral file provided with the later paper, which is available online \cite{femoco_integrals}. All the computational details can be found in Ref. \cite{nike}.

In all three cases, we have employed the Fiedler method \cite{barcza_2011} to order the active space orbitals on a one-dimensional lattice. The ordering optimization was iterated about four times by means of the QC-DMRG calculations with increasing bond dimensions varying from $M=256$ up to $M=1024$, which were followed by the calculations of the single-site entropies and mutual information necessary for the Fiedler method and the warm-up procedure \cite{Szalay-2015a, barcza_2011, Legeza-2003b}. At least three sweeps with the final orbital ordering and actual bond dimensions were performed before measuring the individual timings in the middle of the sweep. All the QC-DMRG calculations were initialized with the CI-DEAS procedure \cite{Szalay-2015a, Legeza-2003b} and they were performed with the \textsf{MOLMPS} program. 

\section{Results and discussion}
\label{section_results}

\begin{figure*}[!ht]
  \subfloat[Davidson procedure \label{kyticka_2048_dav}]{%
    \includegraphics[width=8cm]{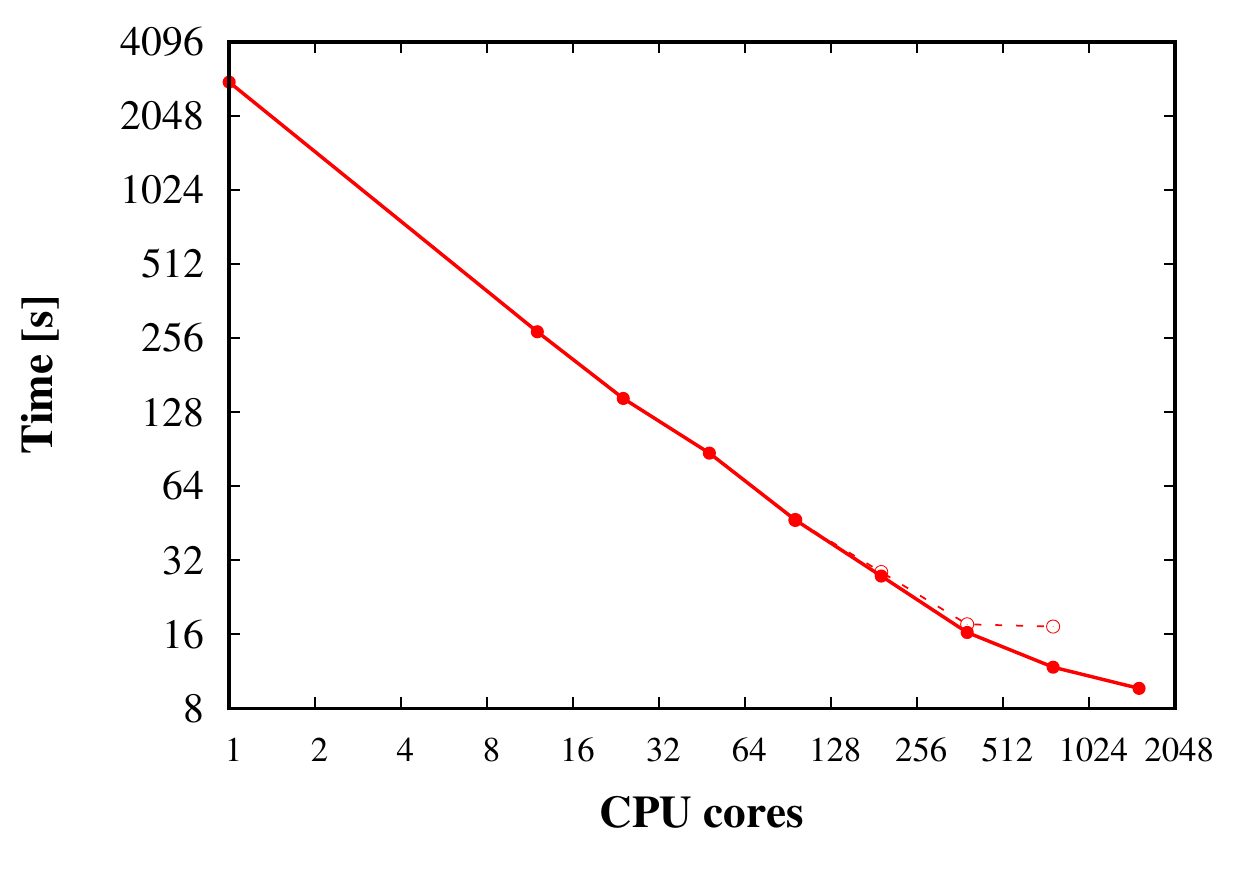}
  }
  \hfill
  \subfloat[Pre-summation and renormalization \label{kyticka_2048}]{%
    \includegraphics[width=8cm]{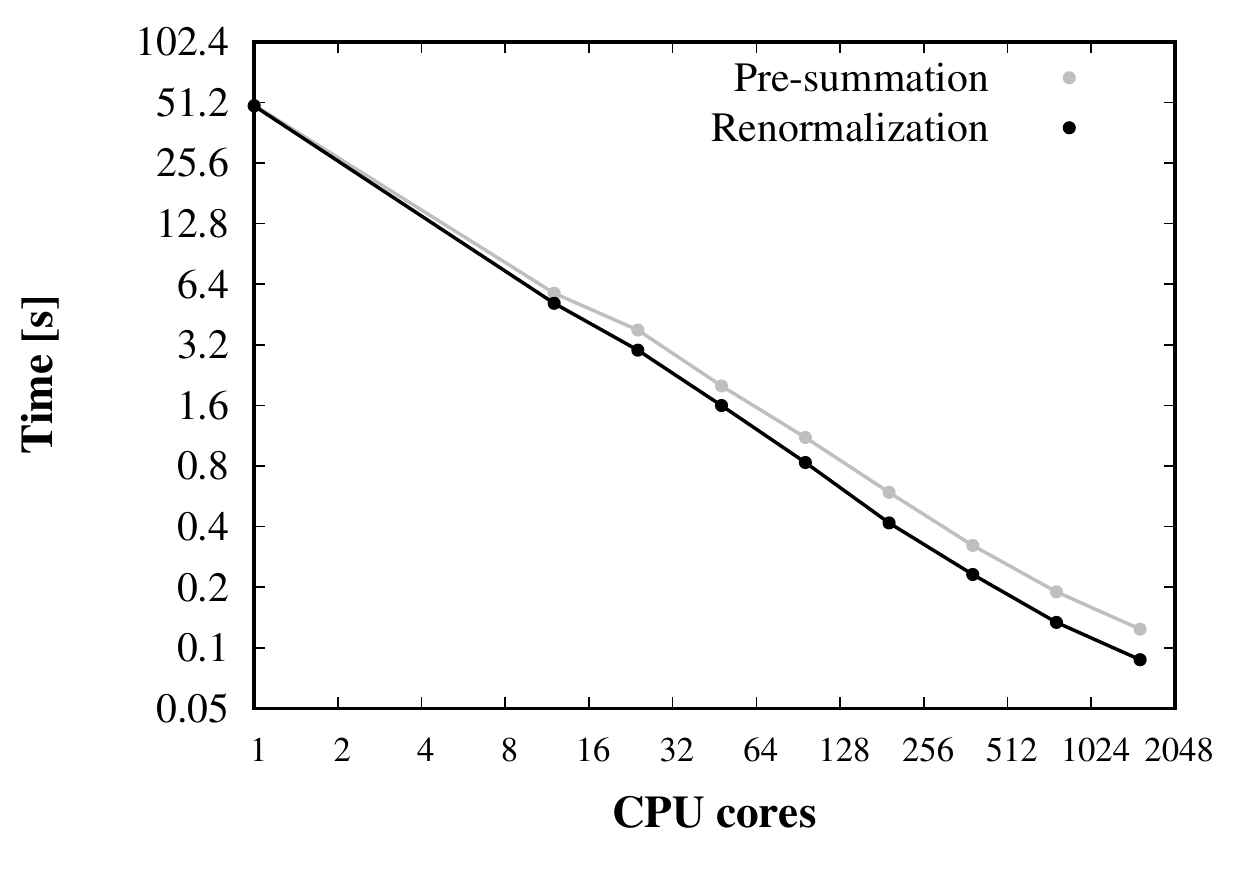}
  }  
  \caption{Timings of the individual parts of one QC-DMRG iteration corresponding to the middle of the sweep performed on the Fe(II)-porphyrin model [CAS(32,34)] with bond dimension $M = 2048$.}
  \label{kyt2k}
\end{figure*}

\begin{figure*}[!ht]
  \subfloat[Davidson procedure \label{kyticka_4096_dav}]{%
    \includegraphics[width=8cm]{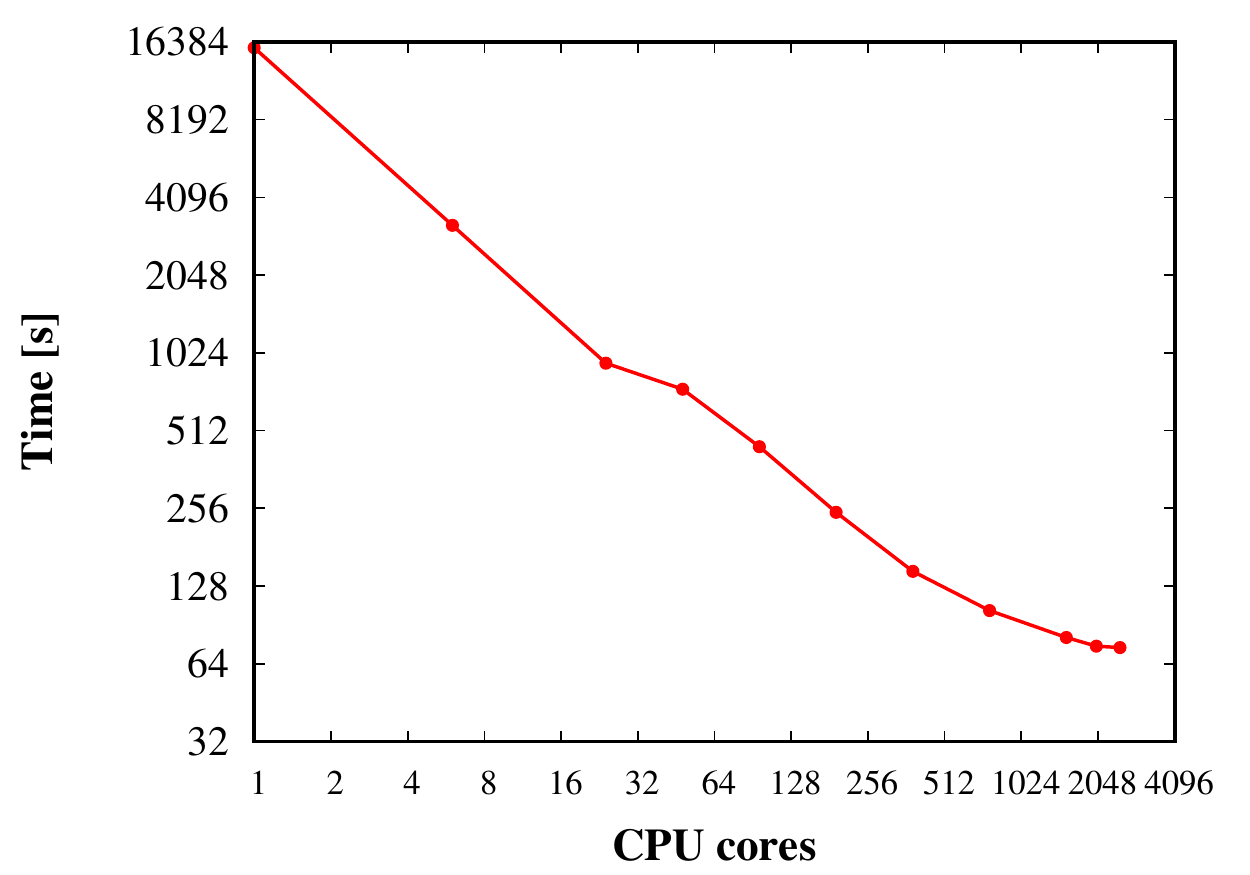}
  }
  \hfill
  \subfloat[Renormalization \label{kyticka_4096}]{%
    \includegraphics[width=8cm]{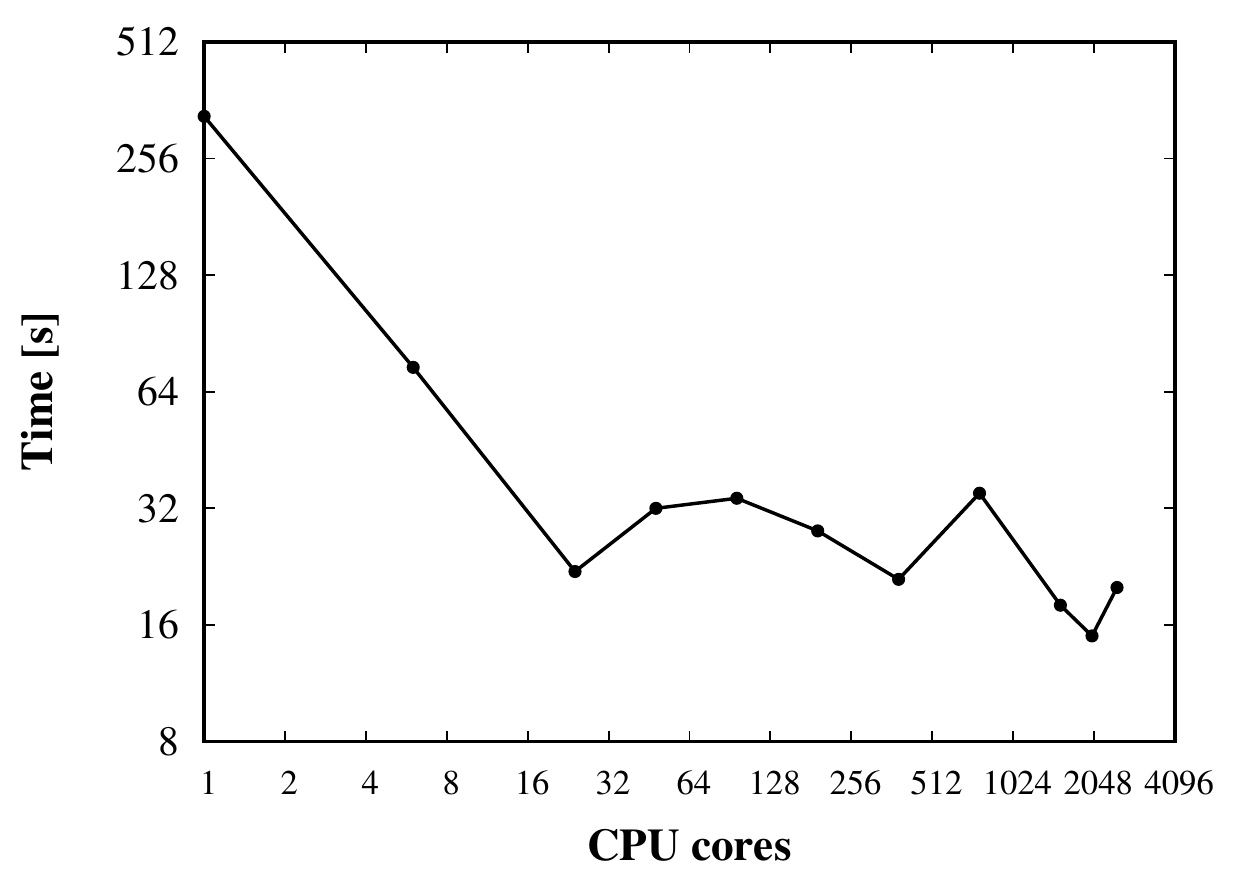}
  }  
  \caption{Timings of the Davidson procedure and the renormalization of the QC-DMRG iteration corresponding to the middle of the sweep performed on the Fe(II)-porphyrin model [CAS(32,34)] with bond dimension $M = 4096$.}
  \label{kyt4k}
\end{figure*}

\begin{figure*}[!ht]
  \subfloat[Davidson procedure \label{kyticka_8192_dav}]{%
    \includegraphics[width=8cm]{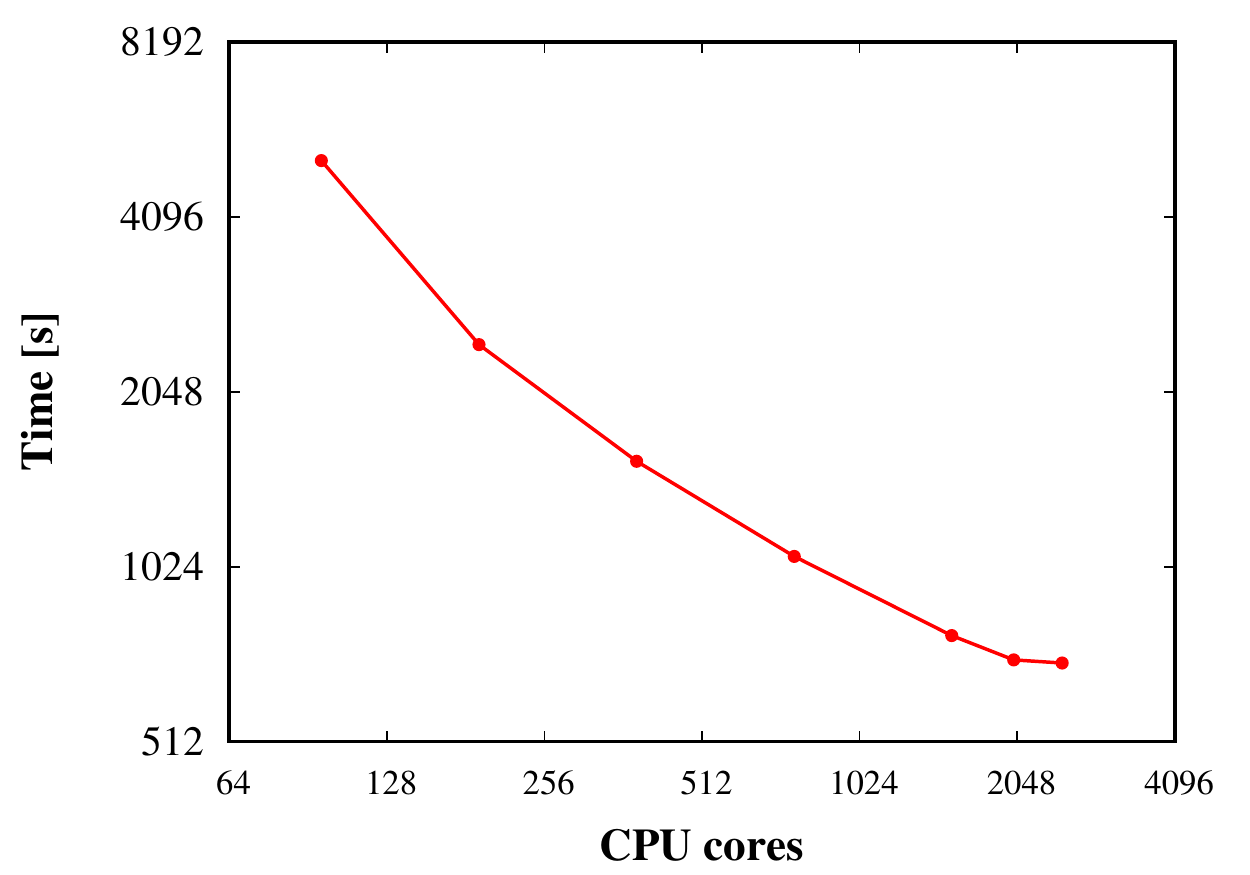}
  }
  \hfill
  \subfloat[Renormalization \label{kyticka_8192}]{%
    \includegraphics[width=8cm]{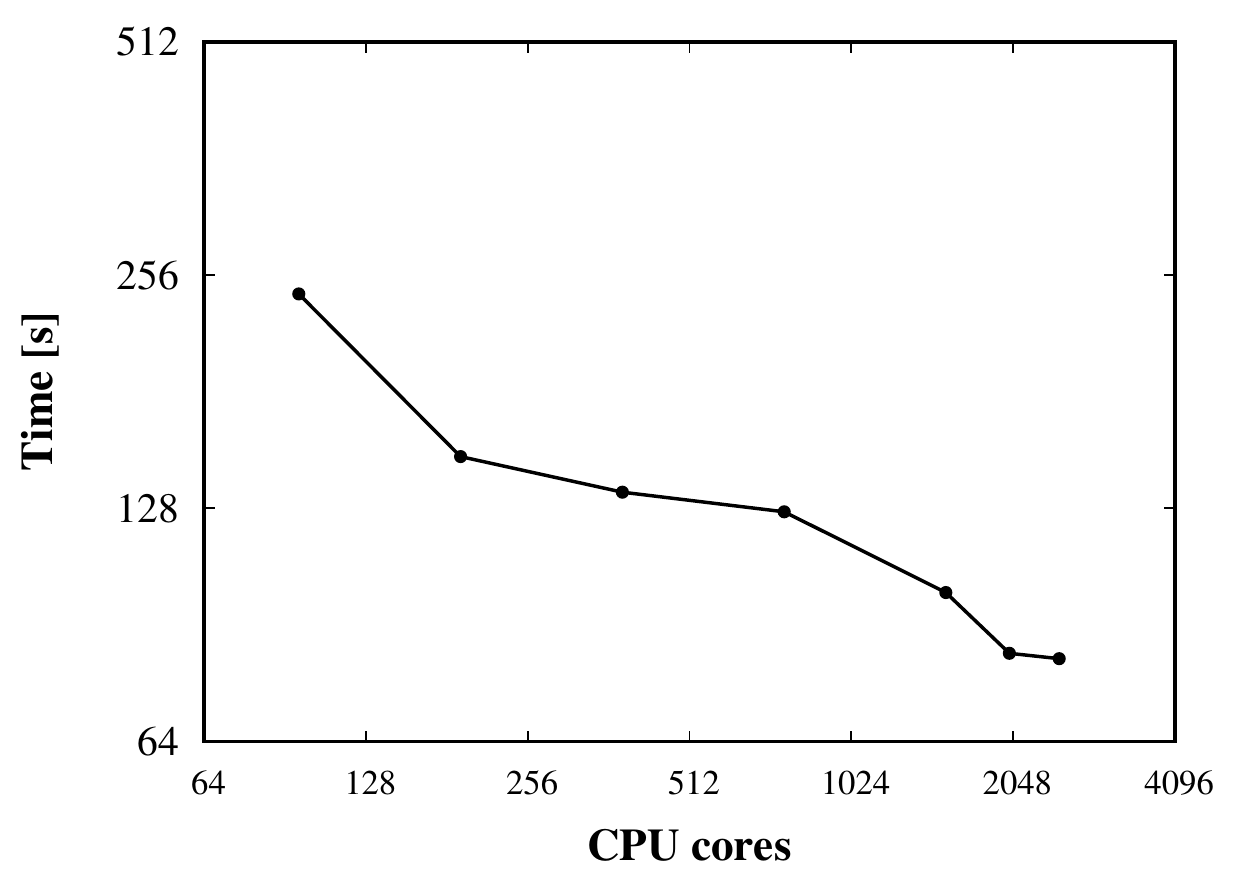}
  }  
  \caption{Timings of the Davidson procedure and the renormalization of the QC-DMRG iteration corresponding to the middle of the sweep performed on the Fe(II)-porphyrin model [CAS(32,34)] with bond dimension $M = 8192$.}
  \label{kyt8k}
\end{figure*}

\begin{figure}[!ht]
  \begin{center}
    \includegraphics[width=8cm]{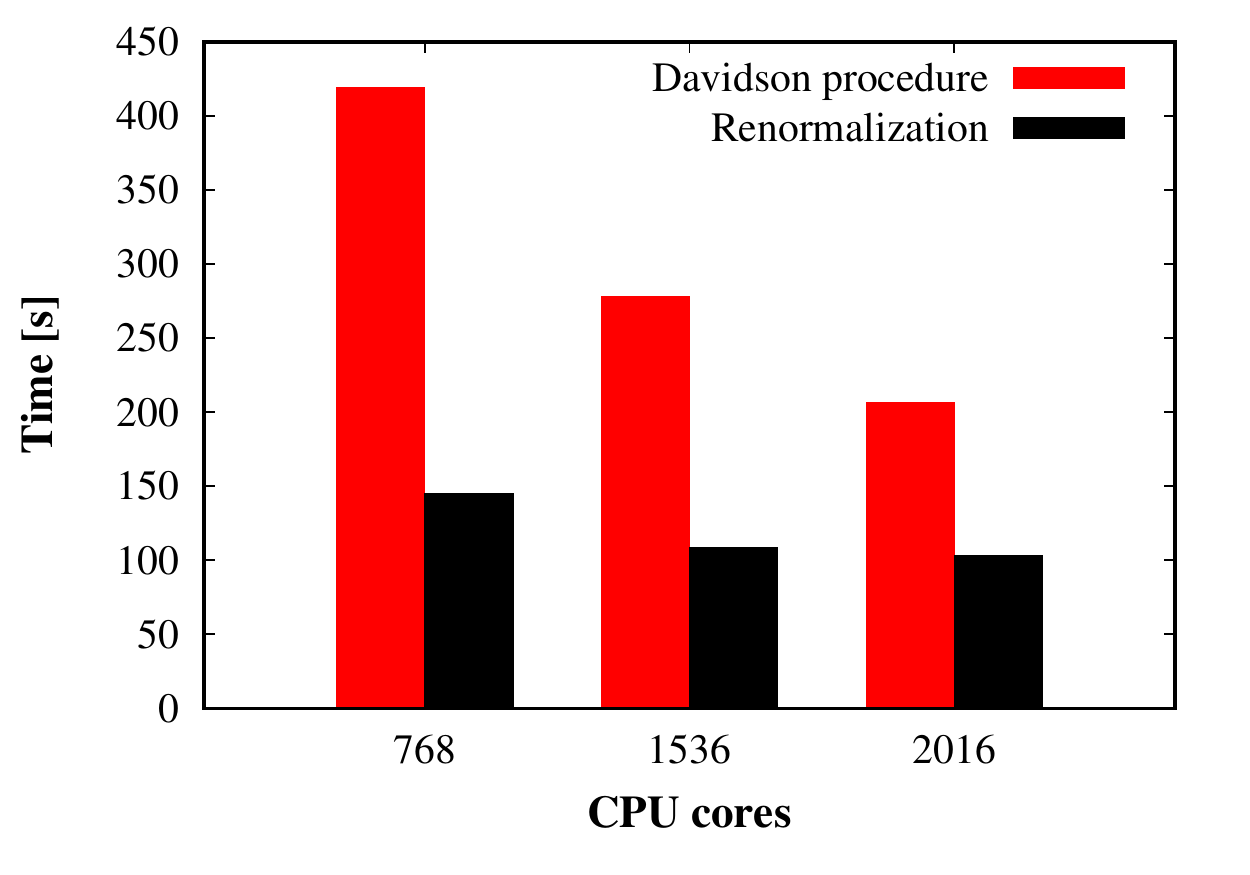}
    \caption{Timings of the Davidson procedure and the renormalization of the QC-DMRG iteration corresponding to the middle of the sweep performed on the defected $\pi$-conjugated anthracene tetramer [CAS(63,63)] with bond dimension $M = 4096$.}
    \label{jelinek_4096}
  \end{center}
\end{figure}

\begin{figure}[!ht]
  \begin{center}
    \includegraphics[width=8cm]{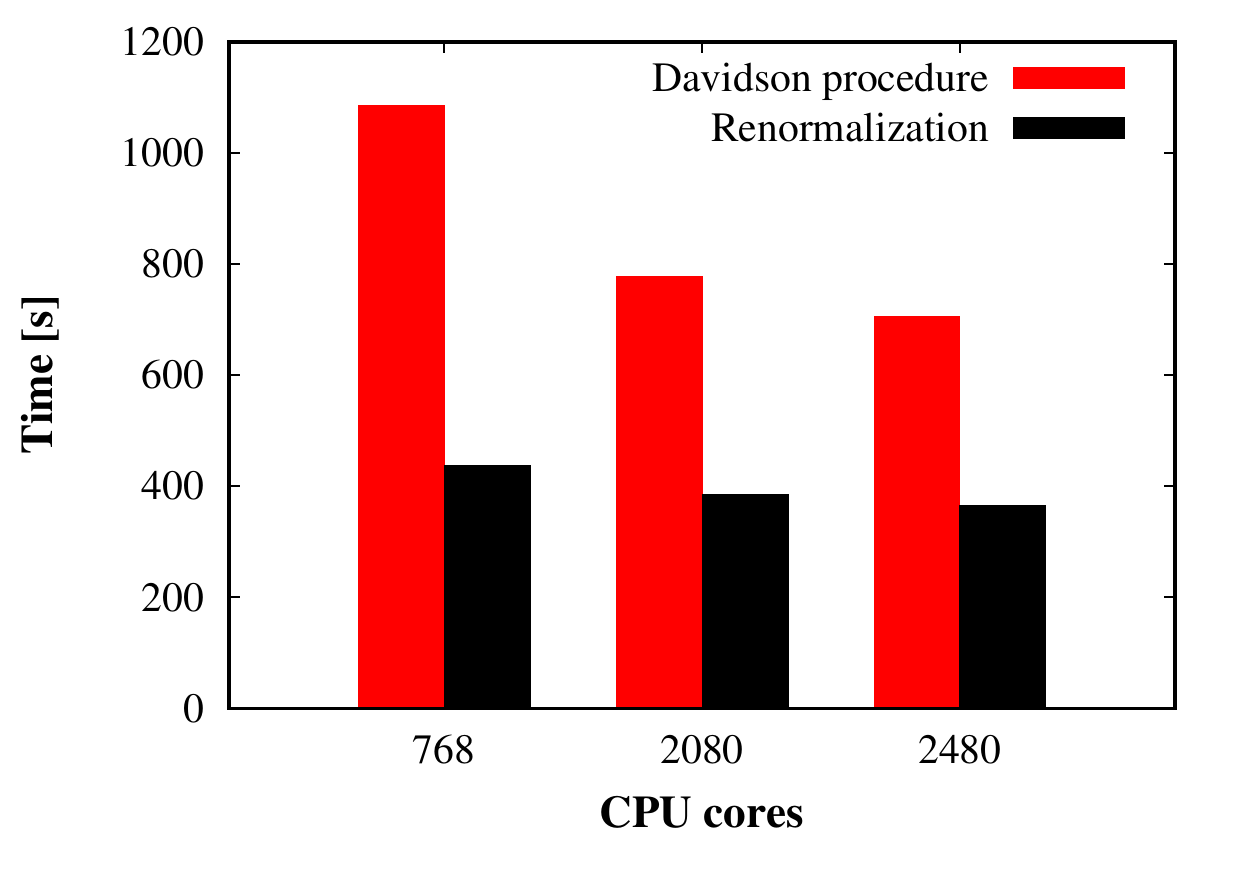}
    \caption{Timings of the Davidson procedure and the renormalization of the QC-DMRG iteration corresponding to the middle of the sweep performed on the FeMoco cluster [CAS(113,76)] with bond dimension $M = 6000$.}
    \label{femoco_6000}
  \end{center}
\end{figure}

The timings of the individual parts of one QC-DMRG iteration for all tested systems corresponding to the middle of the sweep, which is the most time consuming, are summarized in Figures \ref{kyt2k}, \ref{kyt4k}, \ref{kyt8k}, \ref{jelinek_4096}, and \ref{femoco_6000}. In the present study, we were interested solely in the scaling characterictics of our parallel scheme, we therefore do not present any energies (they would be out of context anyway). 
Nevertheless, the detailed study of the Fe(II)-porphyrin model has already been submitted \cite{porph} and  
chemistry-oriented papers about the remaining two systems should also appear soon.

All calculations were performed on the Salomon supercomputer of the Czech national supercomputing center in Ostrava with the following hardware: 24 cores per node (2 x Intel Xeon E5-2680v3, 2.5 GHz), 128 GB RAM per node and InfiniBand FDR56 interconnect. We have used up to 2480 CPU cores for a single QC-DMRG calculation.
We present timings only for the three main parts discussed in the text, namely Hamiltonian diagonalization via Davidson procedure, operator pre-summation, and renormalization. Other parts
including e.g. the formation and diagonalization of the reduced density matrix or broadcastings are marginal for the presented cases.

We have tested the performance of the local memory model on
the example of the smallest system [Fe(II)-porphyrin model, $M = 2048$]. 
As can be seen in Figure \ref{kyticka_2048_dav}, the 
Davidson algorithm scales almost ideally up to approx. 500 CPU cores and still shows a good performance up to approx. 1500 CPU cores. The tiny bump at 48 CPU cores is caused by the fact that despite no communication is needed during the execution of individual tasks, the final result scattered in chunks among nodes has to be gathered (by means of the reduce function) after each Davidson step, which requires a small amount of communication.

The dashed line in Figure \ref{kyticka_2048_dav} around 512 CPU cores corresponds to the same treatment of operator combinations acting on both blocks (left and right) as in case of the GM model, i.e. performing the loop over orbital indices inside tasks. One can see that such a treatment is not suitable for the local memory model where all data are easily accessible locally.

The pre-summation of operators and renormalization in Figure \ref{kyticka_2048} also show almost perfect scaling.

On the example of the Fe(II)-porphyrin model with $M = 4096$, we demonstrate the transition from the local to the GM model. This case still fits into the memory of a single node, however, we have employed the GM approach in order to see the effect of communication. For this and further cases, we do not present scalings for pre-summations since they scale almost perfectly (no need for communication).

The effect of communication on scaling of the Davidson algorithm (Figure \ref{kyticka_4096_dav}) is apparent when going from a single node (24 CPU cores) to 48 CPU cores. The scaling is a lot worse than in case of the local memory model, which is definitely not surprising, but a reasonable improvement can be seen up to aprrox. 1500 cores.
The situation is worse for renormalization (Figure \ref{kyticka_4096}), which scales badly in this case. However, an important point to stress is that it is much faster than the Davidson algorithm itself (75 sec. vs 20 sec. for 2496 CPU cores).

The Fe(II)-porphyrin model with $M = 8192$ in Figure \ref{kyt8k} is 
a memory demanding example which requires at least 4 nodes (of 128 GB). The scaling of the Davidson algorithm (Figure \ref{kyticka_8192_dav}), as well as the renormalization (Figure \ref{kyticka_8192}) is indeed similar to the $M = 4096$ case, despite the fact that the tasks required significantly more intensive communication.

The defected $\pi$-conjugated anthracene tetramer with $M = 4096$ (Figure \ref{jelinek_4096}) and FeMoco cluster with $M = 6000$\footnote{In case of the Davidson algorithm, the block which was not going to be enlarged corresponded to M = 4000. This does not affect the renormalization though.} (Figure \ref{femoco_6000}) represent the most challenging problems which require larger number of nodes and also large amount of communication. Our parallel approach scales up to approx. 2000 CPU cores and in case of the FeMoco cluster, there is still a non-negligible improvement (10\%) up to approx. 2500 CPU cores.
In all the tested cases, the renormalization is adequately faster than the Davidson algorithm.

The performance analysis of the largest calculations mentioned above still shows signs of the inter-node imbalance, which keeps us a room for further improvement and will be the subject of future work.

\section{Conclusions}
\label{section_conclusions}

In this paper, we have presented the first attempt (to our best knowledge) to exploit the supercopmuter platform for QC-DMRG calculations. We have developed the parallel scheme based on the MPI global memory library which combines operator and symmetry sector parallelisms. We have tested its performance on three different molecules with the active spaces ranging from to 34 up to 76 orbitlas and various bond dimensions. For smaller computations (smaller active spaces and bond dimensions) we have achieved almost perfect scaling. For larger calculations, which did not fit into the memory of a single node, we have achieved worse, but still reasonable scaling up to about 2000 CPU cores. Our largest calculation corresponds to the FeMoco cluster [CAS(113,76)] with bond dimension $M=6000$ on 2480 CPU cores.
We believe that further acceleration is possible when the problem of the inter-node imalance is solved in future. As was also discussed in the text, another possible source of speed-up can be GPU units.

In summary, we have shown that the most challenging problems of the current electronic structure may be calculated by means of the QC-DMRG method on a supercomputer in a fraction of time of the few-node calculation.

\section*{Acknowledgment}


We would like to thank Pavel Jel\'{i}nek for providing us with the DFT optimized geometry of the defected $\pi$-conjugated anthracene tetramer.

This work has been supported by the Czech Science Foundation (grant no. 18-18940Y),
the Center for Scalable and Predictive methods for Excitation and Correlated phenomena (SPEC), which is funded by the U.S. Department of Energy (DOE), Office of Science, Office of Basic Energy Sciences, the Division of Chemical Sciences, Geosciences, and Biosciences,
the Hungarian  National  Research,  Development  and  Innovation  Office  (grant  no. K120569),
and  the  Hungarian  Quantum  Technology  National  Excellence  Program  (project no. 2017-1.2.1-NKP-2017-00001).
All the computations were carried out on the Salomon supercomputer in Ostrava, we would therefore like to acknowledge the support by the Czech Ministry of Education, Youth and Sports from the Large Infrastructures for Research, Experimental Development and Innovations project ``IT4Innovations National Supercomputing Center - LM2015070".

\bibliography{references,ors}

\end{document}